\documentclass[11pt]{article}

\usepackage[margin=1in]{geometry}
\usepackage{amsmath,amssymb,amsthm}
\usepackage{mathtools}
\usepackage{graphicx}
\usepackage[scaled=1.0]{helvet}
\usepackage{fancyhdr}
\usepackage{booktabs}
\usepackage{enumitem}
\usepackage{xcolor}
\usepackage[hyphens]{url}
\usepackage[colorlinks=true,linkcolor=blue,citecolor=blue,urlcolor=blue]{hyperref}
\usepackage{cleveref}
\usepackage{braket}		% Dirac notation
\usepackage{tikz}
\usetikzlibrary{arrows.meta,positioning,decorations.pathmorphing}

\newtheorem{definition}{Definition}
\newtheorem{axiom}{Axiom}

\newtheorem{principle}{Principle}

\newcommand{\ts}{t_{\mathrm{s}}}
\newcommand{\Tc}{T_{\mathrm{c}}}
\newcommand{\Tplus}{\mathcal{T}_{+}}
\newcommand{\Tminus}{\mathcal{T}_{-}}
\newcommand{\Rcal}{\mathcal{R}}
\newcommand{\Iplus}{I_{+}}
\newcommand{\Iminus}{I_{-}}

\pdfoutput=1    % Required for arXiv

\begin{document}

\title{\fontfamily{phv}\selectfont{\LARGE{\bfseries{Subtime: Reversible Information Exchange\\ and the Emergence of Classical Time}}}}

\author{
Paul L.\ Borrill\\[4pt]
\textit{D\AE D\AE LUS}\\
\texttt{paul@daedaelus.com}\\
ORCID: 0000-0001-9787-7979
}

\date{\vspace{-10pt}}

\maketitle

\begin{abstract}
We formalize the concept of \emph{subtime}---a reversible mode of information interchange within entangled systems---and show how classical time emerges as an asymptotic limit through decoherence.
Building on the photon clock model, in which a single photon confined between two ideal mirrors creates an alternating causality regime, we develop a process-theoretic formalization using the Oreshkov--Costa--Brukner framework extended with an explicit time-reversal duality condition.
We introduce Perfect Information Feedback (PIF) as the information-theoretic realization of this reversibility, demonstrating that mutual information is conserved in any closed causal loop and that entropy quantifies the degree of unreflected causality.
We define the Reversible Causal Principle (RCP): every causal relation possesses a conjugate dual, and entropy, energy dissipation, and the classical arrow of time appear only when these alternating components decohere or fail to reflect perfectly.
The framework unifies Wheeler--Feynman absorber theory, Bennett's reversible computation, Shannon's communication theory, and the process matrix formalism under a single symmetry principle, and identifies experimentally accessible signatures in reversible digital links and quantum switch experiments.
The arrow of time, in this picture, records the universe's imperfect causal echo.
\end{abstract}

\vspace{10pt}

\vspace{20pt}

%=========================================================
\section{Introduction}
\label{sec:introduction}
%=========================================================

\begin{quotation}
\footnotesize
\noindent ``Church's thesis and the Turing machine are rooted in the concept of \emph{doing one thing at a time}. But we do not really know what `doing' is---or time---without a complete picture of quantum mechanics and the relationship between the still mysterious wave-function and macroscopic observation.''
\begin{flushright} \emph{--- Andrew Hodges,\\ Alan Turing: Life and Legacy of a Great Thinker}~\cite{teuscher2004} \end{flushright}
\normalsize
\end{quotation}

The assumption that information flows forward in time---that causes precede effects, that transmitters precede receivers, that entropy increases monotonically---is so deeply embedded in physics and information theory that it functions not as a tested hypothesis but as an unexamined axiom.
We call this the \emph{Forward-In-Time-Only} (FITO) assumption.
Shannon's Mathematical Theory of Communication~\cite{shannon1948} presupposes it implicitly: information flows from source to destination along a channel parameterized by a monotonically increasing time coordinate.
Eddington's thermodynamic arrow~\cite{eddington1928} codified it for physics.
Lamport's logical clocks~\cite{lamport1978} embedded it in distributed computing.
In each domain, the temporal direction of information flow is treated as given---a fixed feature of reality rather than a convention imposed by the observer.

This paper questions that assumption.

In 2013, the author proposed the concept of \emph{subtime} ($\ts$)---a reversible mode of information interchange within entangled systems---as an alternative to the FITO framework~\cite{borrill2013fqxi,borrill2015springer}.
(That essay appeared alongside Rovelli's relational-information program~\cite{rovelli2015fqxi} in the same Springer volume.)
The core insight was that within an isolated entangled system, information exchange between particles is perpetually reversible: a photon bouncing between two atoms constitutes a ``hot potato protocol'' in which subtime increments along the forward path and decrements on the return, yielding a net-zero temporal displacement.
Classical time ($\Tc$) emerges only as this symmetry breaks through decoherence, producing the monotonic, irreversible accumulation of change that we experience macroscopically.

That original formulation was deliberately non-mathematical---a conceptual argument aimed at reframing the discussion rather than providing formal machinery.
In the intervening years, the physics community has made significant progress on the question of temporal ordering.
Oreshkov, Costa, and Brukner~\cite{oreshkov2012} introduced the process matrix formalism for indefinite causal order.
Experimental realizations of the quantum switch~\cite{procopio2017,rubino2017} demonstrated that superpositions of causal orders are physically realizable.
The two-state vector formalism~\cite{aharonov1964} and Page--Wootters mechanism~\cite{page1983} provided additional frameworks for time-symmetric quantum theory.
And the discovery of opposing arrows of time in open quantum systems~\cite{nature2025opposing} has placed the question of temporal directionality at the center of current research.

This paper formalizes what the 2013 essay proposed conceptually.
We develop the photon clock model---two ideal mirrors confining a photon in an alternating causality regime---as a minimal physical realization of subtime.
We embed this model within the process matrix formalism, extending it with an explicit time-reversal duality condition that distinguishes \emph{alternating} from \emph{indefinite} causal order.
We introduce Perfect Information Feedback (PIF) as the information-theoretic realization of this reversibility, and show that the conservation of mutual information in a closed causal loop is the informational analogue of unitarity.
We then unify these perspectives under the Reversible Causal Principle (RCP), which states that every causal relation possesses a conjugate dual, and that entropy, dissipation, and the classical arrow of time appear only when these alternating components decohere.

The structure of the paper is as follows.
Section~\ref{sec:subtime-classical} defines subtime and classical time and establishes their relationship.
Section~\ref{sec:photon-clock} develops the photon clock model as a minimal physical realization.
Section~\ref{sec:process-formalization} provides the process-theoretic formalization.
Section~\ref{sec:information-conservation} establishes information conservation within subtime.
Section~\ref{sec:emergence} analyzes how classical time emerges through decoherence.
Section~\ref{sec:connections} connects the framework to existing programs in physics.
Section~\ref{sec:experimental} identifies experimental and engineering signatures.
Section~\ref{sec:discussion} discusses implications.

%=========================================================
\section{Subtime and Classical Time}
\label{sec:subtime-classical}
%=========================================================

\subsection{Definitions}

We begin with two foundational definitions that distinguish reversible from irreversible temporal modes.

\begin{definition}[Subtime]
\label{def:subtime}
\emph{Subtime} ($\ts$) is the reversible information interchange within an isolated entangled system.  In subtime, information exchange between constituent particles proceeds bidirectionally: a photon traversal from Alice to Bob is followed by a return from Bob to Alice, such that the net informational displacement is zero.  The system evolves through its configuration space, but every state visited can be revisited by reversal.  Subtime is \emph{propechronus}---local, neighboring, directionless.
\end{definition}

\begin{definition}[Classical Time]
\label{def:classical-time}
\emph{Classical time} ($\Tc$) is the progressive, monotonic, irreversible accumulation of informational change that emerges as an entangled system decoheres into its environment.
Formally, $\Tc$ is the absolute value of the accumulated subtime intervals:
\begin{equation}
\label{eq:Tc}
\Tc = \left| \sum_{i} \ts^{(i)} \right|,
\end{equation}
where $\ts^{(i)}$ are the signed subtime increments along individual traversals.
Because $\Tc$ represents an absolute value, it is always non-negative and monotonically non-decreasing---recovering the familiar property that ``time marches forward.''
\end{definition}

The crucial distinction is this: within an entangled system, information exchange is symmetric and the temporal direction is undefined.
A photon bouncing between two atoms creates no net temporal displacement in $\ts$, because each forward traversal is balanced by a return.
The system explores its configuration space reversibly---what we call the ``hot potato protocol.''
Only when the system interacts with its environment (through decoherence, measurement, or absorption) does the symmetry break, and one direction of information flow becomes irreversible.
This broken symmetry is $\Tc$.

\subsection{The Category Mistake}

The FITO assumption---that information must flow forward in time---conflates two distinct concepts.
In the language of Gilbert Ryle~\cite{ryle1949}, it commits a \emph{category mistake}: it treats a convention of description (the arrow of time in our protocols and diagrams) as a fact about nature (the direction of causation in physical processes).

Consider how deeply this mistake is embedded.
Shannon's channel model~\cite{shannon1948} parameterizes information flow by a time coordinate $t$ that increases monotonically from transmitter to receiver.
Lamport's happens-before relation~\cite{lamport1978} imposes a partial order on events that is, by construction, acyclic.
TCP's three-way handshake assumes that acknowledgments arrive ``after'' transmissions.
In each case, the forward direction of time is not derived from the physics of information exchange---it is imposed by the design of the protocol.

The question this paper addresses is: what happens when we remove this imposition?
What is the natural temporal structure of information exchange, prior to any FITO assumption?
The answer, we argue, is subtime: a bidirectional, reversible mode of information interchange that is the default state of isolated quantum systems, from which classical time emerges as a special (decoherent) case.

\subsection{Retroactive Non-discernability}

A key principle connects these definitions to observable physics.

\begin{principle}[Retroactive Non-discernability]
\label{prin:retroactive}
In an entangled system, a state that has been visited, departed from, and then revisited through reversal of the photon path is \emph{indistinguishable} from the original state.
No measurement performed within $\Tc$ can determine how many times the system has cycled through a given configuration in $\ts$.
\end{principle}

This is the temporal analogue of Boltzmann's indistinguishability of identical particles.
Just as we cannot label individual atoms in a gas, we cannot count the number of traversals a photon has made between two entangled atoms.
The recurrence number $N$ is fundamentally uncountable---not because of practical limitations, but because the forward and reverse traversals are ontologically identical.

The implications are striking.
What we observe in $\Tc$ as a single event---the arrival of a photon at a detector---may correspond to any odd number of traversals in $\ts$.
Our instruments function as a ``quantum stroboscope,'' sampling reality in quick flashes separated by periods of invisible reversible evolution.

%=========================================================
\section{The Photon Clock Model}
\label{sec:photon-clock}
%=========================================================

To give subtime a concrete physical realization, we introduce the \emph{photon clock model}: a single photon confined between two ideal mirrors, representing observers Alice ($A$) and Bob ($B$).

\subsection{Setup}

Consider two perfect mirrors facing each other, with a single photon trapped between them.
Each reflection constitutes a lightlike exchange event.
As long as the photon remains confined, the system exists in an \emph{alternating causality regime}: causal influence oscillates reversibly between $A$ and $B$, without a definite direction of time.

When Alice first encounters the photon, she has no knowledge of how far or for how long it has traveled.
She reflects it perfectly toward Bob, reversing both momentum and phase.
From her standpoint, no local time elapses between emission and reception---the photon's worldline is a single lightlike event with zero proper duration.
When Bob receives and reflects the photon, he too possesses no information about Alice's existence---only that a photon arrived with proper time zero.
Only after multiple exchanges can either observer infer a round-trip duration using their own local clock.
This measured interval defines an emergent geometric separation between $A$ and $B$, effectively reconstructing spacetime distance from the statistics of information exchange~\cite{kuypers2025,surace2024}.

\subsection{The Causal Box}

The confined photon and its two mirrors together form an entangled, time-symmetric system.
As long as the photon remains within, its causal configuration is reversible; no absolute ordering between Alice and Bob exists.
We call this the \emph{causal box}---a region of spacetime in which causal order is internally undefined.

Externally, the causal box projects as a single causal diamond with a definite orientation.
Internally, however, it encodes a superposition of time directions.
The alternating forward--backward structure of influence can be represented geometrically as two joined causal triangles: the forward triangle encodes propagation $A \rightarrow B$, while the inverted triangle encodes the return $B \rightarrow A$.
Together they form a closed loop of causation whose total extent defines one invariant ``tick'' of the photon clock.
The meeting point of the two triangles marks the \emph{causal inversion}---the instant at which the arrow of time reverses inside the system.

\subsection{Symmetry Breaking}

A perturbation---such as absorption, decoherence, or asymmetry---breaks the time-reversal symmetry, collapsing the system into a definite causal order.
The external observer then perceives one of four possible classical configurations, corresponding to distinct, definite arrows of time:

\begin{enumerate}[label=(\roman*)]
\item $A$ emits, $B$ absorbs (forward causal diamond, $A \prec B$),
\item $B$ emits, $A$ absorbs (reversed causal diamond, $B \prec A$),
\item simultaneous emission (spacelike separation),
\item simultaneous absorption (convergent causation).
\end{enumerate}

Which configuration is perceived depends on the boundary conditions of the measurement apparatus---not on any intrinsic property of the photon clock itself.
The arrow of time is selected by the act of opening the causal box.

\subsection{An Axiom of Arrival}

The photon clock model suggests a foundational axiom about the relationship between information and time:

\begin{axiom}[All Knowledge Arrives from the Past]
\label{ax:arrival}
Every piece of information---such as the arrival of a photon---constitutes an event that increases memory.
The ordered accumulation of memory defines time's apparent flow; erasure of memory corresponds to the loss of a segment of the past.
What we and our instruments experience as ``time'' is not an external parameter but the sequential acquisition and storage of information.
\end{axiom}

From single-photon interactions to macroscopic communication frames, temporal order emerges as the reconstruction of history from arrivals.
This axiom makes the FITO assumption explicit as a property of observers, not of the underlying physics.

%=========================================================
\section{Process-Theoretic Formalization}
\label{sec:process-formalization}
%=========================================================

We now embed the photon clock model within the process matrix formalism of quantum causal structure, extending it with an explicit time-reversal duality condition.

\subsection{Background: Indefinite Causal Order}

Oreshkov, Costa, and Brukner (OCB)~\cite{oreshkov2012} introduced the \emph{process matrix} formalism to describe correlations among quantum operations without assuming a fixed global causal order.
Within that framework, two operations $A$ and $B$ can exist in a coherent superposition of causal sequences:
\begin{equation}
A \prec B \quad \text{and} \quad B \prec A.
\end{equation}
The process matrix $W$ generalizes the concept of a quantum channel by encoding all physically admissible correlations consistent with local quantum operations but independent of a global temporal reference frame.
The consistency condition
\begin{equation}
\label{eq:ocb-consistency}
\mathrm{Tr}_{\text{out}}(W) = \mathbb{I}_{\text{in}}
\end{equation}
guarantees normalization and excludes causal paradoxes.

\subsection{Alternating vs.\ Indefinite Causal Order}

Alternating Causality (AC) builds upon the OCB foundation but differs conceptually from indefinite causal order (ICO) in a crucial respect.
Whereas ICO describes a \emph{probabilistic or coherent superposition} of orders, AC represents a \emph{deterministic, time-symmetric oscillation} of causal direction.
Instead of an indeterminate mixture of $A \rightarrow B$ and $B \rightarrow A$, AC describes a periodic, reversible exchange of influence between the two parties---a causal process with a built-in temporal phase.

\begin{center}
\begin{tabular}{@{}lll@{}}
\toprule
\textbf{Concept} & \textbf{ICO} & \textbf{AC} \\
\midrule
Relation between $A,B$ & Superposition of $A{\to}B$ and $B{\to}A$ & Oscillation between $A{\to}B$ and $B{\to}A$ \\
Temporal model & Non-factorizable process matrix & Periodic reversible process tensor \\
Reversibility & Implicit via linearity & Explicit via phase alternation \\
Information flow & Coherent mixture of directions & Deterministic bidirectional feedback \\
\bottomrule
\end{tabular}
\end{center}

\subsection{The Time-Reversal Duality Condition}

In the alternating model, we extend the OCB consistency condition to a \emph{two-way consistency constraint}:
\begin{equation}
\label{eq:two-way}
\mathrm{Tr}_{A_{\text{out}}}(W_{AB}) = \rho_{B_{\text{in}}},
\qquad
\mathrm{Tr}_{B_{\text{out}}}(W_{BA}) = \rho_{A_{\text{in}}},
\end{equation}
together with the symmetry condition
\begin{equation}
\label{eq:time-reversal}
\boxed{W_{BA}(t) = W_{AB}(-t)^{\dagger},}
\end{equation}
which enforces a strict time-reversal duality between the forward and reverse causal phases.

Equation~\eqref{eq:time-reversal} is the central mathematical statement of this paper.
It says that the process matrix governing the causal influence from $B$ to $A$ at time $t$ is the Hermitian conjugate of the process matrix governing influence from $A$ to $B$ at time $-t$.
This is not a superposition of orders; it is a \emph{deterministic oscillation} between conjugate causal phases.

\subsection{Process Tensor Decomposition}

The \emph{process tensor} formalism~\cite{milz2021} generalizes the process matrix to extended temporal sequences:
\begin{equation}
\mathcal{T}_{k:0} \in
\mathcal{L}\!\left(
\bigotimes_{i=0}^{k-1}
\mathcal{H}_{i}^{\mathrm{in}} \otimes \mathcal{H}_{i}^{\mathrm{out}}
\right),
\end{equation}
capturing temporal correlations and memory across multiple interventions.

Within Alternating Causality, this tensor naturally decomposes into forward and reverse components:
\begin{equation}
\label{eq:tensor-decomp}
\mathcal{T}_{k:0} = \Tplus + \Tminus,
\qquad
\Tminus(t) = \Tplus(-t)^{\dagger}.
\end{equation}
This time-symmetric structure guarantees full reversibility: each causal phase is the temporal mirror of its counterpart.
The alternating phase relation defines a reversible cycle of causal exchange rather than a probabilistic mixture of directions.

\subsection{The Causal Resonator}

The process-theoretic perspective reveals a geometric interpretation.
In ICO, causal direction is \emph{undefined}---the underlying structure exists in a superposition of orders.
In AC, causal direction is \emph{phase-linked}: it alternates coherently, like a standing wave exchanging energy between two reflectors.
Each causal link thus behaves as a \emph{causal resonator}, sustaining equilibrium between forward and backward propagation.

This directly connects to the photon clock model of Section~\ref{sec:photon-clock}: a photon confined between two mirrors embodies a reversible, alternating causal process whose external measurement collapses to a single direction of time.
The process matrix $W$ encodes the internal dynamics of the causal box; the time-reversal condition~\eqref{eq:time-reversal} ensures that these dynamics are exactly reversible; and the collapse to a definite causal order upon measurement corresponds to the projection of the full process tensor onto a single temporal phase.

The duality condition~\eqref{eq:time-reversal} also mirrors the reversibility of quantum logic operations:
\begin{equation}
U^{-1} = U^{\dagger}.
\end{equation}
AC therefore unifies two paradigms: the quantum-informational notion of causal symmetry and the computational requirement of logical reversibility.
In this sense, AC can be regarded as the \emph{causal analogue of a unitary gate}: time evolution preserves information and admits a well-defined inverse.

%=========================================================
\section{Information Conservation in Subtime}
\label{sec:information-conservation}
%=========================================================

We now translate Alternating Causality into the language of information theory, showing that reversible causal interactions correspond to a form of \emph{Perfect Information Feedback} (PIF).

\subsection{The FITO Channel and Its Generalization}

In Shannon's framework, information flows one way along a channel parameterized by time.
Let $X$ and $Y$ denote the random variables associated with sender and receiver states.
The mutual information is:
\begin{equation}
I(X;Y) = H(X) - H(X|Y),
\end{equation}
where $H$ denotes Shannon entropy.
This describes a FITO channel: information flows from $X$ to $Y$, period.

In a PIF channel, we generalize to bidirectional flow.
Each direction participates symmetrically:
\begin{equation}
\Iplus(X;Y) = \Iminus(Y;X), \qquad I_{\mathrm{total}} = \Iplus + \Iminus.
\end{equation}
The system enforces a conservation of mutual information:
\begin{equation}
\label{eq:info-conservation}
\boxed{\frac{d\Iplus}{dt} + \frac{d\Iminus}{dt} = 0,}
\end{equation}
so that no net informational entropy is produced.
This is the information-theoretic expression of the same reversibility condition that appeared in the process-matrix form~\eqref{eq:time-reversal}.

In a PIF channel, each transmission is both a measurement and a reflection of state---a self-consistent two-way information flow that eliminates the distinction between ``send'' and ``receive'' within a single causal cycle.

\subsection{Entropy Balance and Symmetric Capacity}

The entropy balance of a PIF link can be written:
\begin{equation}
H_{\mathrm{in}} = H_{\mathrm{out}},
\end{equation}
subject to the constraint that every bit transmitted is verified by its mirror:
\begin{equation}
p(x,y) = p(y,x).
\end{equation}
In the Shannon sense, the channel capacity for such a reversible link doubles the degrees of freedom available for encoding:
\begin{equation}
C_{\mathrm{PIF}} = 2\, C_{\mathrm{one\text{-}way}},
\end{equation}
because forward and backward propagating modes each carry distinct, yet phase-locked, informational content.
Unlike redundancy or forward error correction overhead, this duality does not waste bandwidth; it enforces consistency through symmetry.

\subsection{Landauer Cost Avoidance}

A crucial consequence of information conservation in subtime concerns the thermodynamic cost of computation.
Landauer~\cite{landauer1961} demonstrated that the erasure of one bit of information necessarily dissipates at least $k_B T \ln 2$ of energy as heat.
Bennett~\cite{bennett1973} showed that logically reversible computation avoids this cost entirely, because no information is ever destroyed.

In a PIF system operating in the alternating causality regime, the entropy budget is strictly zero under ideal operation: all information exchange is balanced by reflection.
The Landauer cost of erasure is avoided because no information is ever destroyed; each bit's epistemic complement remains stored in its mirror state.
This provides a physically grounded mechanism for reversible computation at the network level, and connects the abstract mathematics of Section~\ref{sec:process-formalization} to concrete thermodynamic predictions.

\subsection{Entropy as Unreflected Causality}

The information-theoretic picture provides a new interpretation of entropy.
In the standard thermodynamic account, entropy measures disorder or the number of accessible microstates.
In the subtime framework, entropy measures something more specific: the degree to which causal reflections are incomplete.

When a PIF channel operates perfectly---every forward transmission matched by a backward echo---entropy production is zero.
When the echo is imperfect (due to decoherence, absorption, or boundary asymmetry), some fraction of the transmitted information fails to be reflected.
This unreflected information constitutes entropy production:
\begin{equation}
\label{eq:entropy-unreflected}
\Delta S = I_{\mathrm{transmitted}} - I_{\mathrm{reflected}}.
\end{equation}
Entropy, in this view, quantifies the ``openness'' of causal loops---the failure of the universe to perfectly echo its own information.

%=========================================================
\section{The Emergence of Classical Time}
\label{sec:emergence}
%=========================================================

\subsection{The Reversible Causal Principle}

We can now state the organizing symmetry of the entire framework.

\begin{principle}[Reversible Causal Principle (RCP)]
\label{prin:rcp}
A process satisfies the Reversible Causal Principle if every causal relation $A \leftrightarrow B$ possesses a conjugate dual such that
\begin{equation}
\label{eq:rcp-local}
\mathcal{P}_{BA}(t) = \mathcal{P}_{AB}(-t)^{\dagger},
\end{equation}
and the composite process over any closed causal loop obeys
\begin{equation}
\label{eq:rcp-global}
\oint_{\Gamma} d\phi = 0.
\end{equation}
The first condition expresses \emph{local reversibility}; the second, \emph{global causal neutrality}.
\end{principle}

All preceding constructions can be recast in a common operator form:
\begin{equation}
\label{eq:unified-operator}
\Rcal(t) = \Tplus(t) + \Tminus(-t)^{\dagger},
\end{equation}
where $\Tplus$ and $\Tminus$ represent forward and reverse evolutions.
The invariant quantity
\begin{equation}
\label{eq:info-invariant}
\langle\psi|\Rcal^{\dagger}\Rcal|\psi\rangle = \mathrm{const.}
\end{equation}
encodes conservation of total information, analogous to probability conservation in unitary quantum mechanics.
Entropy growth corresponds to a deviation from unitarity of $\Rcal$.

The RCP extends Noether's logic: just as time-translation symmetry implies energy conservation, \emph{causal symmetry implies information conservation}.

\subsection{How Time's Arrow Emerges}

Within RCP, the arrow of time is \emph{emergent}.
It appears when alternating causal phases decohere or when boundary conditions block perfect reflection:
\begin{equation}
\label{eq:decoherence-break}
\Tminus(-t)^{\dagger} \neq \Tplus(t),
\end{equation}
introducing a small imaginary component in the action:
\begin{equation}
S \rightarrow S + i\epsilon,
\end{equation}
which breaks time-reversal symmetry and produces effective entropy.
The macroscopic direction of time thus records the universe's imperfect causal echo.

This mechanism has a natural interpretation in the photon clock model.
When the mirrors are perfect and the photon remains confined, $\Tminus(-t)^{\dagger} = \Tplus(t)$ exactly, and the system inhabits subtime.
When a perturbation allows the photon to escape---by absorption into one mirror, or by coupling to environmental degrees of freedom---the equality fails.
The escaped information becomes irretrievable, and the system ``falls into'' classical time.

\subsection{The Decoherence Cascade}

The transition from subtime to classical time is not abrupt but hierarchical.
Consider a chain of $n$ atoms, each entangled with its neighbors through photon exchange.

For the simplest case of two atoms ($n = 2$), the system is almost perfectly reversible: the photon bounces between them with near-unit fidelity, and the recurrence time is short.
As we add atoms, the number of discernible configurations grows combinatorially, and the probability of returning to any previously visited state decreases.
By the time $n$ reaches mesoscopic scales ($n \sim 10^{6}$--$10^{12}$), the recurrence time exceeds the age of the universe, and the system is effectively irreversible---not because the laws have changed, but because the phase space is too vast for the causal echoes to find their way back.

This is the emergence of $\Tc$ from $\ts$: not a phase transition, but an asymptotic limit.
Classical time is what subtime looks like when you can no longer keep track of the reversals.

\subsection{Correspondence Table}

The framework achieves a unification across four domains:

\begin{center}
\begin{tabular}{@{}lll@{}}
\toprule
\textbf{Domain} & \textbf{Symmetric Quantity} & \textbf{Irreversibility Indicator} \\
\midrule
Quantum / Process Matrix & $W_{BA} = W_{AB}^{\dagger}$ & Non-Hermiticity \\
Information Theory & $\Iplus = \Iminus$ & Mutual-information loss \\
Electrodynamics & $A^{\mathrm{ret}} + A^{\mathrm{adv}}$ & Radiation damping \\
Computation & $U^{-1} = U^{\dagger}$ & Landauer erasure \\
\bottomrule
\end{tabular}
\end{center}

Across all domains, reversibility manifests as the vanishing of an asymmetry term---informational, energetic, or computational.
The RCP provides a single grammar joining Shannon's communication symmetry, Wheeler--Feynman absorber dynamics, and Bennett's reversible computation.

%=========================================================
\section{Connection to Existing Programs}
\label{sec:connections}
%=========================================================

\subsection{Wheeler--Feynman Absorber Theory}

The most direct physical analogue of the subtime framework is the time-symmetric electrodynamics of Wheeler and Feynman~\cite{wheeler1945,wheeler1949}, which removes self-action by requiring that every emission be balanced by an advanced response from future absorbers.
The total electromagnetic potential is the symmetric combination:
\begin{equation}
A_{\mu}^{\mathrm{sym}}(x) = \tfrac{1}{2}\!\left[A_{\mu}^{\mathrm{ret}}(x) + A_{\mu}^{\mathrm{adv}}(x)\right].
\end{equation}
Every radiative act is therefore intrinsically alternating in temporal direction.

In absorber theory, emission occurs only if future absorbers exist to return an advanced echo---a global consistency condition that parallels the feedback symmetry in PIF.
We can make the mapping explicit:
\begin{align}
\text{Forward phase } (A \to B) &\;\leftrightarrow\; A_{\mu}^{\mathrm{ret}}, \\
\text{Reverse phase } (B \to A) &\;\leftrightarrow\; A_{\mu}^{\mathrm{adv}}.
\end{align}
When both directions remain phase-locked, the system forms a standing wave of causality with zero net temporal bias.
Breaking this symmetry---by suppressing one direction---produces apparent irreversibility, exactly as incomplete absorption yields radiation damping in Wheeler--Feynman theory.

The global closure condition in absorber theory (the universe as a complete absorber) corresponds directly to the RCP's global causal neutrality condition~\eqref{eq:rcp-global}.
Entropy therefore quantifies the degree of unreflected information---the failure of the ``cosmic absorber'' to return its advanced echoes.

\subsection{The Two-State Vector Formalism}

Aharonov, Bergmann, and Lebowitz~\cite{aharonov1964} introduced the two-state vector formalism (TSVF), in which quantum systems are described jointly by a forward-evolving state $|\psi\rangle$ (prepared at the initial time) and a backward-evolving state $\langle\phi|$ (selected at the final time).
Measurement outcomes are determined by both vectors simultaneously.

The subtime framework resonates strongly with the TSVF.
In the photon clock model, the forward traversal $A \rightarrow B$ corresponds to the forward-evolving state, and the return $B \rightarrow A$ to the backward-evolving state.
The alternating causality regime is the physical realization of a system in which both temporal directions are equally real---neither is privileged until measurement selects one.

The key difference is that the TSVF typically treats the backward-evolving state as a mathematical tool for computing weak values and retrodiction, whereas the subtime framework treats it as \emph{ontologically co-equal} with the forward state.
In subtime, the photon really does bounce back; the return path is not a computational fiction but a physical process.

\subsection{Page--Wootters and Relational Time}

Page and Wootters~\cite{page1983} proposed that time can be defined relationally, using one subsystem as a clock for another, within a globally static quantum state.
The universe as a whole satisfies the Wheeler--DeWitt equation $\hat{H}|\Psi\rangle = 0$---it is ``timeless''---while temporal evolution emerges for subsystems conditioned on the clock's reading.

The subtime framework is compatible with the Page--Wootters picture.
An entangled system in the alternating causality regime is internally ``timeless'' in exactly this sense: the photon clock ticks, but no net temporal displacement accumulates.
Classical time ($\Tc$) emerges when subsystems decohere and begin to serve as effective clocks for one another.
The photon clock model thus provides a microscopic mechanism for the Page--Wootters emergence: the ``clock'' is the photon bouncing between mirrors, and ``time'' is the accumulation of irreversible interactions with the environment.

\subsection{Rovelli's Relational Quantum Mechanics}

Rovelli~\cite{rovelli1996,rovelli2015fqxi,rovelli2018} argues that quantum states are not absolute properties of systems but relations between systems.
There is no ``view from nowhere''---every measurement is relative to an observer.
This is precisely the local-observer perspective that the subtime framework adopts.
In subtime, there is no global time coordinate; each entangled pair has its own local temporal experience ($\ts$), and $\Tc$ emerges only through the web of decoherent interactions between pairs.

%=========================================================
\section{Experimental and Engineering Signatures}
\label{sec:experimental}
%=========================================================

The subtime framework is not merely a philosophical reinterpretation.
It makes predictions that are, in principle, distinguishable from standard quantum mechanics and, in some cases, already realized in engineering practice.

\subsection{Reversible Digital Links}

The most immediate testbed is \emph{Open Atomic Ethernet} (OAE)~\cite{borrill2025oae}, a protocol that implements bidirectional, semantically reversible communication at the physical layer.
In OAE, each 8-byte slice transmitted by the sender is immediately echoed by the receiver's SERDES interface, forming a discrete causal reflection:
\begin{equation}
\text{Send: } s(t), \quad \text{Receive: } r(t) = s(-t)^{*}.
\end{equation}
A mismatch between sent and received slices corresponds to phase decoherence---a detectable, local, reversible error.
Because both directions are active simultaneously, a failure of one leg is instantly visible as a loss of symmetry rather than an undetected forward error.

The measurable signature of RCP in this setting is the suppression of informational entropy: a perfectly operating OAE link should exhibit zero net entropy production across the bidirectional channel, with entropy appearing only at the boundaries where the link couples to irreversible (FITO) protocols.

The bisynchronous FIFO---a silicon-proven component that enables bidirectional state transfer without temporal ordering~\cite{borrill2025bsfifo}---provides hardware evidence that FITO is not physically required at the link layer.

\subsection{Quantum Switch Experiments}

Experimental realizations of the quantum switch~\cite{procopio2017,rubino2017} demonstrate superpositions of causal orders.
The subtime framework predicts that these experiments should be reinterpretable in terms of alternating (rather than indefinite) causal order.
Specifically, if the control qubit is prepared in a state that selects a deterministic alternation between $A \rightarrow B$ and $B \rightarrow A$ (rather than a coherent superposition), the observable statistics should be distinguishable from standard ICO predictions in the presence of decoherence.

The key prediction is that an alternating causal protocol should exhibit \emph{suppressed entropy growth} compared to a superposition-based protocol, because the deterministic phase structure of AC preserves information more efficiently than the stochastic phase structure of ICO.

\subsection{Optical and Superconducting Cavities}

High-finesse optical cavities~\cite{kimble1998} and superconducting microwave resonators~\cite{blais2021} already realize physical systems in which photons are confined between reflecting boundaries for extended periods.
These systems naturally implement the photon clock model, with the cavity mirrors playing the role of Alice and Bob.

The subtime prediction is that the internal dynamics of such cavities---prior to photon loss or absorption---should exhibit exact time-reversal symmetry at the level of individual quantum trajectories.
While this is already implicit in quantum optics, the subtime framework provides a specific information-theoretic metric: the conservation of mutual information across the bidirectional field modes inside the cavity.

%=========================================================
\section{Discussion}
\label{sec:discussion}
%=========================================================

\subsection{What This Paper Claims}

We have formalized the concept of subtime---the reversible information interchange within entangled systems---and shown that:

\begin{enumerate}[label=(\arabic*)]
\item The photon clock model provides a minimal physical realization of subtime, in which a single photon confined between two mirrors creates an alternating causality regime with no definite direction of time.

\item The process matrix formalism, extended with the time-reversal duality condition $W_{BA}(t) = W_{AB}(-t)^{\dagger}$, provides the mathematical structure for alternating (as opposed to indefinite) causal order.

\item Perfect Information Feedback (PIF) is the information-theoretic realization: mutual information is conserved in any closed causal loop, and entropy quantifies unreflected causality.

\item The Reversible Causal Principle unifies these perspectives: every causal relation possesses a conjugate dual, and the classical arrow of time emerges when decoherence prevents perfect causal reflection.

\item The framework is experimentally accessible through reversible digital links, quantum switch experiments, and cavity QED systems.
\end{enumerate}

\subsection{What This Paper Does Not Claim}

We do not claim that subtime replaces standard quantum mechanics.
The mathematical structure we develop is consistent with unitary quantum theory; it provides a reinterpretation of temporal structure within that theory, not a modification of it.
Nor do we claim that the arrow of time is illusory---it is real, in the same sense that temperature is real: it is an emergent, macroscopic property that accurately describes the behavior of systems far from the reversible limit.

What we do claim is that the FITO assumption---the idea that information \emph{must} flow forward in time---is a category mistake.
It confuses a property of our protocols and measurements (the direction in which we record information) with a property of the underlying physics (the direction of causal influence).
The subtime framework makes this distinction precise and shows that the apparent forward flow of time is a special case of a more symmetric structure.

\subsection{Implications for the Arrow of Time Debate}

The subtime framework contributes to the ongoing debate about the origin of time's arrow~\cite{price1996,barbour1999,rovelli2018,smolin2013} by offering a specific mechanism for its emergence.
Rather than postulating a low-entropy initial condition (the Past Hypothesis) or treating the arrow as a fundamental asymmetry of the laws, the framework locates the arrow in the \emph{failure of causal reflection}---the inability of the universe to perfectly echo its own information at macroscopic scales.

This shifts the question from ``why does time have a direction?'' to ``why is the universe not a perfect absorber?''---a question that connects naturally to cosmological boundary conditions and the thermodynamics of decoherence.

\subsection{Causal Structure as Gauge}

A suggestive implication of the RCP is that causal structure may have the character of a gauge field.
Choosing a temporal orientation---deciding which direction is ``future''---is a local gauge transformation in a causal fiber bundle.
The physical content lies in the gauge-invariant quantities: the conservation laws (information, energy) and the curvature (entropy production).
This perspective aligns with recent work on causal structure in quantum gravity~\cite{hardy2007,oreshkov2012} and suggests that the subtime framework may find natural expression within the broader program of quantum causal structure.

%=========================================================
\section*{Acknowledgments}
%=========================================================

The author thanks Sahas Munamala, David Johnston, and Dugan Hammock for their contributions to the Alternating Causality framework;
Robert Metcalfe and David Boggs for foundational discussions on Ethernet architecture;
and the Open Compute Project Timecard/TAP working group for providing a forum in which these ideas were first presented.
The original subtime concept was developed during the Time One project and awarded 4th prize in the 2013 FQXi essay contest ``It From Bit, or Bit From It?''
AI writing assistance (Anthropic Claude) was used in the preparation of this manuscript; the author takes full responsibility for all scientific content.

%=========================================================
% BIBLIOGRAPHY
%=========================================================

\end{document}